\documentclass[aps,pra,superscriptaddress]{revtex4}
\usepackage{graphicx}

\usepackage{amsfonts}
\usepackage{amsmath}
\usepackage{color}
\usepackage{bm}
\newcommand{\bsy}[1]{\ensuremath{\boldsymbol{#1}}}

\begin{document}

\title{Production of optical phase space vortices with non-locally distributed mode converters}

\author{R. M. Gomes}
 \affiliation{Instituto de F\'{\i}sica,
Universidade Federal do Rio de Janeiro, Caixa Postal 68528, Rio de
Janeiro, RJ 21941-972, Brazil}
\author{A. Salles}
\affiliation{Niels Bohr Institute, Blegdamsvej 17, 2100 Copenhagen, Denmark}
\author{F. Toscano}
 \affiliation{Instituto de F\'{\i}sica, Universidade Federal do Rio
              de Janeiro, Caixa Postal 68528, Rio de Janeiro, RJ 21941-972,
              Brazil}
\author{P. H. Souto Ribeiro}
 \affiliation{Instituto de F\'{\i}sica,
Universidade Federal do Rio de Janeiro, Caixa Postal 68528, Rio de
Janeiro, RJ 21941-972, Brazil}
\author{S. P. Walborn}
\affiliation{Instituto de F\'{\i}sica, Universidade Federal do Rio
de Janeiro, Caixa Postal 68528, Rio de Janeiro, RJ 21941-972,
Brazil}

\begin{abstract}
Optical vortices have been observed in a wide variety of optical systems. They can be observed directly in the wavefront of optical beams, or in the correlations between pairs of entangled photons. We present a novel optical vortex which appears in a non-local plane of the two-photon phase space, composed of a single degree of freedom of each photon of an entangled pair.  The preparation of this vortex can be viewed as a ``non-local" or distributed mode converter.  We show how these novel optical vortices of arbitrary order can be prepared in the spatial degrees of freedom of entangled photons.   
\end{abstract}

\maketitle

\section{Introduction}
Optical vortices arise from phase singularities, typically associated to the orbital angular momentum (OAM) of light.  The OAM of light beams is a classical feature that has only been studied in detail recently \cite{allen03}.
Among other applications \cite{grier03}, a key motivation for this study is related to applications in quantum information\cite{molina07}. 
These potential applications have motivated the study of the OAM in the context
of the spatial correlations between photon pairs produced in spontaneous parametric down conversion(SPDC) \cite{arnaut01,franke-arnold02,aiello05b}. 
Experimentally, entanglement in the OAM of photon pairs has been observed \cite{mair01,vaziri03a,walborn04a,langford04,altman05,oemrawsingh05,leach09} and quantified in recent experiments \cite{peeters07,pires10}.  These correlations are a consequence of the conservation of the OAM in the SPDC process, which arises from the transfer of the angular spectrum of the pump beam to the wave function of the down-converted photons \cite{monken98a,torres05}.  
\par
Investigations of OAM correlations between down-converted photons generally rely on one of two types of detection systems.  In the first type, holographic masks and single mode optical fibers are used to project the down-converted fields onto modes with well-defined OAM \cite{mair01}, such as the Laguerre-Gaussian (LG) modes.  In other experiments, point-like detectors are scanned in the transverse detection plane, and the characteristic doughnut modes are observed in the coincidence count distributions \cite{walborn04a,altman05}.  In Refs. \cite{walborn04a,altman05}, one detector was kept fixed while the other was scanned in the two-dimensional detection plane.  The intensity distribution of a doughnut mode does not appear in the single-photon counts but rather only in the joint distribution of coincidence counts. The correlations observed with these two types of detection systems have been shown to be characteristic of entanglement.         
\par
A different kind of non-local optical vortex has been recently observed \cite{gomes09a}. Here the term ``non-local" is related to the  two-dimensional space in which the optical vortex appears.  A doughnut-shaped coincidence distribution and phase dislocation was observed, but the pump beam is not prepared in an LG mode and neither of the twin photons are projected onto states with well-defined OAM.  In all previous experiments, OAM correlations were observed only if either the pump, or at least one of the down-converted photons is prepared or projected onto a state with OAM different from zero. 
This is not the case in Ref. \cite{gomes09a}, where the optical vortex appears in the coordinate plane composed by the position variable of photon 1 and the wave vector variable of photon 2.  
\par
  In this paper, we generalize the scheme employed in Ref. \cite{gomes09a} and show that the presence of the optical vortex is due to the implementation of a non local mode converter. We demonstrate that it is possible to create non-local optical vortices of arbitrary order through manipulation of the pump laser beam.  In addition, we show that this can be done separately and simultaneously in two spatial dimensions.  
\section{Non-local Optical Vortex}
\begin{figure}
\begin{center}
\includegraphics[width=12cm]{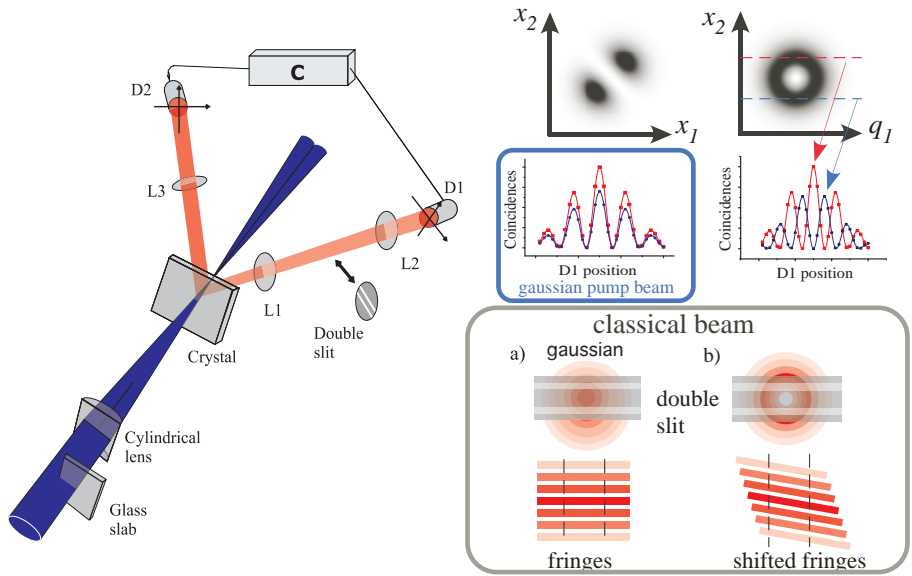}
\end{center}
\caption{Sketch of the experimental set-up (left) and results (right) for the observation of a non-local optical vortex \cite{gomes09a}.  The top right illustrates the coincidence distributions obtained in the $x_1,x_2$ and $q_1,x_2$ planes, as well as the double slit interference obtained in coincidence counts.  The blue and red curves correspond to the interference patterns obtained with different positions of detector 2.  In the case of a HG pump beam, a shift is observed, while in the case of a Gaussian pump beam, no shift is observed (blue box).  For comparison, the grey box illustrates double slit interference with a a) Gaussian and b) Laguerre-Gaussian beam $LG_{0}^{1}$.}
\label{setup}
\end{figure}
 Fig. \ref{setup} shows a sketch of the experimental set-up used in Ref.\cite{gomes09a} to produce a nonlocal optical vortex.   The pump laser is prepared in a first order Hermite-Gaussian mode (HG) $HG_{10}$, which is achieved by passing half of the  Gaussian laser beam through a 
thin glass slide so that a phase difference of $\pi$ is produced between the
two halves of the pump beam. Further propagation
in free space provides spatial filtering necessary to clean up the beam, producing a good
quality $HG_{10}$ mode.   This beam is then focused onto a nonlinear crystal with a cylindrical lens and produces entangled photons through spontaneous parametric 
down-conversion (SPDC). Twin photons with the same wavelength are collected and detected through narrow band 
interference filters.   For photon 1, an imaging system is used to map the source distribution  ($x,y$) onto the detection
plane.  For photon 2, a Fourier lens maps the wave vector ($q_x,q_y$) distribution of the source onto the detection plane.  
Both detectors are scanned in a line along the vertical (either $x$ or $q_x$) detection axis, giving a two-dimensional non-local distribution of coincidence counts. 
A typical distribution is sketched in Fig. \ref{setup}, showing the doughnut
shape that is an indication of the presence of an optical vortex.  For comparison, Fig. \ref{setup} also sketches the coincidence distribution when both photons are measured in the position basis $x$. We see that in this case the coincidence distribution mimics the pump beam 
inside the crystal \cite{monken98a}.  It is important to notice that the pump beam is not prepared
in an LG mode and the down-converted photons are not projected onto LG modes. 
\par
The simple observation of a doughnut shape is not enough to prove the existence of the optical vortex. It is also 
necessary to obtain information about the azimuthal phase distribution. In Ref.\cite{gomes09a},
this was done through a double slit interference experiment.   For a classical light beam, the phase singularity associated to an LG beam can be evidenced by observing a shift in the interference fringes \cite{sztul06}, as illustrated in the inset of Fig. \ref{setup}.  
This method was used to evidence the phase dependence of the nonlocal vortex in Ref. \cite{gomes09a}.  As shown in Fig. \ref{setup}, a double slit aperture is inserted into the path of photon 1 and the detector is scanned
to register the interference fringes in the coincidence distribution for two different positions of detector 2.   The measurements are performed for two configurations of the pump beam.  First pumping with a zero order Gaussian beam and second with the  
$HG_{10}$ mode.  In the first case, the sets of observed fringes are in phase, while in the second the observed fringes are out of phase (see figure \ref{setup}), as in the case of a classical $LG_{0}^{1}$ beam.
\par
We will show in section \ref{sec:nonlocalmodconv} that this experiment can be viewed as the implementation of mode conversion between a nonlocal HG and LG mode.  To do so, let us briefly review mode conversion of HG and LG beams.     
\section{Mode-Conversion Between Hermite-Gaussian and Laguerre-Gaussian Modes}
In classical optics,  the Hermite-Gaussian (HG) and Laguerre-Gaussian (LG)  beams are solutions to the
paraxial Helmholtz equation \cite{saleh91} in cartesian and cylindrical coordinates, respectively.    
We will write the HG modes as $HG_{nm}=HG_n(x)HG_m(y)$, 
where $n$ ($m$) is the number of zeroes of the Hermite polynomials $H_n(x)$ 
($H_m(y)$).  The order of the HG modes is $n+m$.  The LG modes can be written as $LG_p^\ell(x,y)$, where the order is $|\ell|+2p$.   
  It is well known that the
Laguerre-Gaussian beams carry orbital angular momentum of $\ell \hbar$ per photon \cite{allen03}.  
\par
 Beijersbergen {\it et al.} showed how one can exploit the Guoy phase of a paraxial beam to convert HG modes into LG modes or vice-versa \cite{beijersbergen93}. 
 The LG modes, for example,  can be described as a superposition of HG modes of the same order. A simple example is the mode $LG^{1}_{0}$, which is given by the superposition of $HG_{10}$ and $HG_{01}$ with a relative phase of $\pi/2$:
$LG^{1}_{0} = \frac{1}{\sqrt{2}}(HG_{10} + iHG_{01})$. 
At the same time, the {\it diagonal} HG mode $DHG$ is defined as 
\begin{equation}
DHG_{nm}(x,y) = HG_{nm}\left(\frac{x+y}{\sqrt{2}},\frac{x-y}{\sqrt{2}} \right)
\label{eq:dhgef}
\end{equation}
 and $DHG_{01}$ is given by $DHG_{01} = \frac{1}{\sqrt{2}}(HG_{10} + HG_{01})$. 
One can see that the $DHG_{01}$ mode can be converted into the $LG^{1}_{0}$ mode  by introducing a relative phase between the HG components.  This relative phase can be introduced using a mode converter, which is a set of cylindrical lenses which produces an astigmatic region \cite{beijersbergen93}, giving rise to the relative phase shift.   
\par
 In general, the DHG modes are given by \cite{beijersbergen93}
 \begin{equation}
 DHG_{nm}(x,y)=\sum_{j=0}^{N=n+m}b(n,m,j)HG_{N-j,j}(x,y), 
 \label{eq:DHGgen}
 \end{equation}
while the higher order modes $LG^{\ell}_{p}$ are given by
\begin{equation}
LG^{\ell}_{p}(x,y)=\sum_{j=0}^{N=n+m}i^{j}b(n,m,j)HG_{N-j,j}(x,y),
\label{eq:LGgen}
\end{equation}
with $\ell=n-m$, $p=\min(n,m)$ and $b(n,m,j)=\left [\frac{(N-j)!j!}{2^{N}n!m!}\right]^{1/2}\frac{1}{j}\frac{d^{j}}{dt^{j}}[(1-t)^{n}(1+t)^{m}]|_{t=0}$.
For the higher-order beams, each component $HG_{N-j,j}$ in expansion (\ref{eq:DHGgen}) picks up a phase $i^j$ upon propagation through the astigmatic region of the mode converter \cite{beijersbergen93}.  In this way, one can transform a general $DHG_{nm}$ mode into an $LG^{\ell}_{p}$ mode of the same order.    
 
\section{Non-local Conversion of Higher-order Modes in Two Dimensions}
\label{sec:nonlocalmodconv}
In this section, we will show a general method of mode conversion using entangled photons, such as those produced in SPDC.  This method is a ``non-local" or distributed mode converter, since the $LG$ mode is generated in a non-local phase space of the spatial degrees of freedom of the entangled  photons.  This generalizes our previous result reported in Ref. \cite{gomes09a} to the case of higher-order modes and two spatial dimensions.      
\par
A fundamental characteristic of the SPDC process is that the transverse profile of the pump field is transmitted to the two photons state \cite{monken98a}. In the monochromatic and paraxial approximations, the two photon state  produced by SPDC can be written as 
\begin{equation}
|{\psi}\rangle = \int d\bsy{q}_1 \int  d\bsy{q}_2 \Psi(\bsy{q}_1,\bsy{q}_2) |\bsy{q}_1\rangle|\bsy{q}_2\rangle,
\end{equation}
where $\bsy{q}_j$ is the transverse wave vector of photon $j$, and $|\bsy{q}_j\rangle$ represents a single-photon state with well-defined frequency and polarization.  The wave function in wave vector space is given by
\begin{equation}
\Psi(\bsy{q}_{1},\bsy{q}_{2})=v(\bsy{q}_{1}+\bsy{q}_{2})\gamma(\bsy{q}_{1}-\bsy{q}_{2}),
 \label{SPDC}
\end{equation}
where $v$ is the angular spectrum of the pump beam and $\gamma$ is the phase matching function \cite{monken98a}. Let us assume that the down-converted photons are degenerate and quasi-collinear, so that the phase matching function simplifies to \cite{torres05}:  $\gamma(\bsy{q})\propto \mathrm{sinc}[\lambda L(q_{x}^2+q_{y}^2)/8 \pi]$, 
where $\lambda$ is the
wavelength of the pump beam and $L$ is the length of the crystal.  
Using a Gaussian spatial filter on the down-converter photons described in Ref.\cite{gomes09a}
we can tailor the phase matching function
$\gamma({\bf q})$ so it can be well approximated by a Gaussian:  $\gamma(\bsy{q}) \approx HG_{00}(\bsy{q},\lambda,w)$, where the waist is given by $w \approx 8 \pi/ \lambda L$.  Let us further assume that the pump beam is prepared in an HG mode $v(\bsy{q}) = HG_{nm}(\bsy{q},\lambda,w_0)$, which is characterized by the wavelength $\lambda$ and waist $w_0$ of the pump beam.  In wave vector representation, the HG beams are
 \begin{align}
HG_{nm}(q_{x},q_{y}) = D_{nm}H_{n}\left(\frac{w_0 q_{x}}{\sqrt{2}}\right)H_{m}\left(\frac{w_0 q_{y}}{\sqrt{2}}\right) 
 e^{\left(-\frac{w_0^2(q_{x}^2+q_{y}^2)}{4}
\right)}  e^{\left[-i
(n+m+1)\eta(z)\right]},
\end{align}
where $w_0$ is the beam waist, $H_n$ is a Hermite polynomial and $D_{nm}$ is a constant \cite{walborn05b}. Mathematically, the HG beam with wavelength $\lambda$ and waist $w_0$ can be rewritten as  $HG_{nm}(\bsy{q},\lambda,w_0) = HG_{nm}(\bsy{q}/\sqrt{2},2\lambda,\sqrt{2}w_0)$.  
In this case the wave function is given by
\begin{equation}
\Psi(\bsy{q}_{1},\bsy{q}_{2}) = HG_{nm}\left(\frac{q_{x_{1}}+q_{x_{2}}}{\sqrt{2}},\frac{q_{y_{1}}+q_{y_{2}}}{\sqrt{2}}\right) HG_{00}\left(\frac{q_{x_{1}}-q_{x_{2}}}{\sqrt{2}},\frac{q_{y_{1}}-q_{y_{2}}}{\sqrt{2}}\right).
\label{spdc}
\end{equation}
From here on, all HG and LG functions are characterized by the wavelength of the down-converted photons $\lambda_c=2 \lambda$ and waist $w_c = \sqrt{2} w_0$.  
Using $HG_{nm}(x,y)=HG_n(x)HG_m(y)$ we can rewrite 
\begin{equation}
\Psi(\bsy{q}_{1},\bsy{q}_{2}) =
HG_{n0}\left(\frac{q_{x1}+q_{x2}}{\sqrt{2}},\frac{q_{x1}-q_{x2}}{\sqrt{2}}\right)HG_{m0}\left(\frac{q_{y1}+q_{y2}}{\sqrt{2}},\frac{q_{y1}-q_{y2}}{\sqrt{2}}\right).                     
 \label{eq:DHG}
\end{equation}
By the definition of the DHG modes (\ref{eq:dhgef}), we recognize that wave function (\ref{eq:DHG}) is equivalent to
\begin{align}
\Psi(\bsy{q}_{1},\bsy{q}_{2}) = & 
DHG_{n0}\left(q_{x1},q_{x2}\right)DHG_{m0}\left(q_{y1},q_{y2}\right) \nonumber \\
= & \sum_{j=0}^{n}b(n,0,j)HG_{n-j,j}(q_{x1},q_{x2})  \sum_{t}^{m}b(m,0,t)HG_{m-t,t}(q_{y1},q_{y2})\nonumber\\
=& \sum_{j=0}^{n}b(n,0,j)HG_{n-j}(q_{x1}) HG_{j}(q_{x2})  
\sum_{t}^{m}b(m,0,t)HG_{m-t}(q_{y1})HG_{t}(q_{y2}).                     
 \label{DHG2}
\end{align}
Wave function (\ref{DHG2}) is a product of DHG modes which appear in non-local coordinate planes composed of one spatial dimension of each down-converted photon.  
Using expansions (\ref{eq:DHGgen}) and (\ref{eq:LGgen}), we see that introducing relative phases of $i^n$ and $i^m$ in the $x$ and $y$ spatial dimensions of photon 2, these DHG modes can be converted into LG modes. 
These relative phases can be introduced by a Fourier transform operation on photon 2
as was done in Ref. \cite{gomes09a}. 
Indeed, as the Hermite-Gaussian functions are eigenfunctions of the Fourier transform $\mathcal{F}$,
{\it i.e.} ${\cal F}[HG_n]=i^nHG_n$, the two-photon wave function is transformed to
\begin{align}
\Phi(\bsy{q}_{1},\bsy{\rho}_{2}) =  
& \sum_{j=0}^{n}i^j b(n,0,j)HG_{n-j,j}(q_{x1},x_{2})  
\sum_{t=0}^{m} i^tb(m,0,t)HG_{m-t,t}(q_{y1},y_{2}) \nonumber \\
= & \, LG^{n}_{0}\left(q_{x1},x_{2}\right)LG^{m}_{0}\left(q_{y1},y_{2}\right).                     
 \label{eq:LG}
\end{align}
Thus, performing a bidimensional Fourier transform on photon 1 implements the required relative phases.   This operation can be viewed as a nonlocal mode converter, where the astigmatism is introduced between the spatial degrees of freedom of photons 1 and 2, rather than the $x$ and $y$ coordinates of a single classical beam.        

\section{Conclusion}

We have presented a method to produce nonlocal optical vortices using entangled photon pairs produced in parametric down-conversion.  The vortex appears in a 
coordinate plane composed of one spatial degree of freedom of one photon
and  the corresponding transverse wave vector degree of freedom of the other photon.We have shown that this kind
of vortex is obtained from a nonlocal or ``distributed" mode converter in which a single lens is placed in each of the down-converted photons.  Pumping a nonlinear crystal with an appropriate $HG_{nm}$ beam, produces a pair of nonlocal optical vortices of order $n$ and $m$ in different spatial directions.  
\begin{acknowledgements}
This work was performed with support from the funding agencies CNPq, CAPES, FAPERJ,  the Brazilian National Institute for Science and Technology of Quantum Information, (INCT-IQ), and EU STREP COQUIT under FET-Open grant number 233747.
 \end{acknowledgements}

\end{document}